\documentclass[manuscript]{acmart}
\usepackage{amsmath}
\usepackage{mathtools}

\usepackage{array}
\usepackage{simplebnf}
\usepackage{url}
\usepackage{qcircuit}
\usepackage{bbold}
\usepackage[cache=false]{minted}
\newcommand{\ket}[1]{\vert #1 \rangle}
\newcommand{\bra}[1]{\langle #1 \vert}

\usepackage{xcolor}

\newcommand{\namesystem}{PiQuant}
\setminted{
	breaklines=true,
	breakanywhere=true,
	breaksymbolleft=\tiny\textcolor{gray}{\ensuremath{\hookrightarrow}}
}

\title{\namesystem: Modeling Quantum Computing with $\pi$-calculus}

\author{Catalin Toma}
\email{catalint747@gmail.com}
\affiliation{%
  \institution{Constructor University}
  \city{Bremen}
  \country{Germany}
}
\author{Manuel Oriol}
\email{manuel.oriol@heig-vd.ch}
\orcid{0000-0003-4069-7626}
\affiliation{%
  \institution{REDS, School of Engineering and Management Vaud, HES-SO}
  \city{Yverdon-les-Bains}
  \country{Switzerland}
}
\affiliation{%
  \institution{Constructor University}
  \city{Bremen}
  \country{Germany}
}

\usepackage{url}
\usepackage{graphicx}

\begin{document}

\begin{abstract}
Quantum computing is a difficult concept to grasp for computer scientists because it is full of mathematical formulas that have no intuitive meaning related to the problem that the code solves.
While these formulae define the way the system behaves, the operational semantics is very far from regular programming languages that reason with variables, assignments, and passing by value the results of computation.
This article takes a step back to define a theoretical framework where quantum programs are not treated as sequential programs, but rather as concurrent programs where variables advertise their states through their own channels and code that uses them is reading that state (and possibly collapsing it). 

The \namesystem~ framework is a simple extension of the $\pi$-calculus that contains additional syntax and semantics for the handling of quantum variables.
\end{abstract}
\maketitle  
\subsection*{Keywords}
lambda calculus, qubits, p-bits


\section{Introduction}\label{sec:introduction}


Quantum computing has been an interesting topic for scientists in various fields since Manin and Feynman claimed some potential quantum advantage~\cite{feynman_qc, manin_qc}.
In 2007, D-Wave launched its 28 qubit quantum computer, the interest in the field has spiked.
Today quantum computers exist and are publicly available. Programmers can test their software on real, functioning quantum hardware. 

The main issue hindering acceptance, is that most programming languages or libraries still focus on manipulating directly qubits to code algorithms. 
This is fine when only a few qubits are available, but in the coming years access to machines with 10000 qubits or more should be available and these approaches will not scale.
To solve this problem, the quantum community needs programming language that allow for higher levels of abstraction.
Some issues that prevent software engineers from understanding how to use quantum computing are superposition and entanglement.
How to make them understandable within a programming language?

This article answers this question by using making a parallel between quantum systems and concurrent systems:
\begin{itemize}
    \item A quantum variable is a channel on which two values are written at the same time with some distribution (based on a density matrix). This tackles superposition.
    \item A set of quantum variables are multiple channels where two values are written at the same time on each channel. This tackles entanglement.
    \item When reading on a channel, these states collapse and we added a choice operator between the potential values based on the density matrix.
\end{itemize}

This semantics better captures the intuition of what happens in a quantum system.
We use general variables as the basis of our reasoning, such variables could be Qubits or more evolved types.

The present work focuses on developing a $\pi$-calculus \cite{pierce, semantics} for modeling quantum programs called PiQuant. 
Significant inspiration for this was taken from the work of Beno\^it Valiron~\cite{benoit_thesis, benoit_url}, although there is less focus here on mathematical details related to category theory or intuitionistic logic. 
The purpose is to ultimately create a tool that is easy for developers to use, defined by solid theoretical foundations.

To achieve this, we depart from the low-level implementations of quantum programming and define a model of computation better adapted to intuitively capture the functioning of quantum programs.
We also present an implementation of a simulator for the language.

Section~\ref{sec:description} presents an informal description of the language. Section~\ref{sec:related} describes state of the art.
Section~\ref{sec:results} shows the typing rules and semantics of the proposed lambda calculus. 
Section~\ref{sec:ex_prog} contains a few programs written in this lambda calculus. Section~\ref{sec:implementation} describes the implementation. 
Finally, section~\ref{sec:conc} presents the conclusions.

\section{Informal description}\label{sec:description}
The main specificity of working with quantum bits (qubits) is that they potentially have an internal state where several values are superposed.
It is only after the first read, which collapses these values, that we know what value is returned by subsequent reads. 
The probability to obtain {\bf 0} or {\bf 1} for a specific qubit is defined by an internal state that can be manipulated by quantum gates.
Such quantum gates can also combine qubits to obtain entangled states for several qubits.
A theoretical framework trying to model such behaviors must then contain these two elements: 
(1) the reading mechanism that collapses, (2) a way to manipulate the quantum state.

Whereas some variations of the $\lambda$-calculus~\cite{ZORZI_2016, Selinger_Valiron_2009, van_Tonder_2004} contain such mechanisms, the nature of passing variables by values is well-suited for modeling quantum variables. 
Our idea is to use the $\pi$-calculus and its read and write matching mechanisms.
Schematically, a superposed qubit $x$ can hold eitehr one of two values $y_1$ and $y_2$ chosen according to a density matrix $\rho$:

\begin{equation*}
    \overline{x}\langle y_1, y_2, \rho \rangle
\end{equation*}

A read operation is encoded as a normal read in the $\pi$-calculus:
\begin{equation*}
    x(y).P
\end{equation*}

When putting them together, it is possible to reduce them and obtain  two processes: one that repeatedly writes the collapsed value and one that has already read it and replaced its value internally (in this case the first value is chosen):

\begin{equation*}
    \overline{x}\langle y_1, y_2, \rho \rangle | x(y).P \rightarrow ... chosing{\ }y_1... \rightarrow (!\overline{x} \langle y_1 \rangle)|x(y).P 
\end{equation*}

Once the resulting classical value any other process would read the same value as P:

\begin{equation*}
    \begin{split}
        \overline{x}\langle y_1, y_2, \rho \rangle | x(y).P | x(z).Q &\rightarrow ... chosing{\ }y_1... \\
        &\rightarrow ((!\overline{x} \langle y_1 \rangle)|x(y).P) | x(z).Q\\
        &\rightarrow (!\overline{x} \langle y_1 \rangle)| P[y_1/y] | x(z).Q\\
        &\rightarrow (!\overline{x} \langle y_1 \rangle)| P[y_1/y] | Q[y_1/z]
    \end{split}
\end{equation*}

The probability to obtain $y_1$ or $y_2$ is calculated using the density matrix $\rho$ in the standard way (take the measurement operator to be $M=|0\rangle \langle 0|$, then the probability to obtain 0, i.e. $y_1$ is $tr(M \rho)$, while the probability to obtain 1, i.e $y_2$). 

That way, the read value is broadcast using the known operator from the $\pi$ calculus. A reduction corresponds to a measurement in quantum computing. As such, the read process is non-deterministic in nature


Multi-qubit quantum states are represented similarly as:
\begin{equation*}
    \overline{x_0 x_1 \ldots x_n}\langle y_{01}, y_{02},\ldots y_{n1}, y_{n2}, \rho \rangle
\end{equation*}
This corresponds to a multi-qubit state with density matrix $\rho$. The values $y_{i1}$ and $y_{i2}$ represent choices encoded in each qubit. Classically these are the values 0 and 1, but one can choose to encode a choice between different values, for example channel names. 

Manipulating the state of qubits can then also be done through similar reduction rules:


\[\Qcircuit @C=1 em @R=1 em {
 \lstick{q_0=\ket{0}} & \ustick{|} \qw& \ctrl{1} &  \ustick{.} \qw& \meter & \ustick{.} \cw&\cw\\
 \lstick{q_1=\ket{0}} & \ustick{|} \qw& \targ  &\ustick{.} \qw& \meter & \ustick{.} \cw&\cw
}\]

In this example we are applying the CNOT gate to the $\ket{00}$ state. Written in PiQuant, this is:
\begin{equation*}
    \begin{split}
        & \overline{x_0 x_1}\langle0,1,0,1, \rho \rangle| CNOT[x_0,x_1].x_0(x_0).x_1(x_1).P \\
        & \rightarrow \overline{x_0 x_1} \langle 0,1,0,1, \rho' \rangle|x_0(x_0).x_1(x_1).P\\
        & \rightarrow \overline{x_1}\langle 0,1, \rho'_{-0}\rangle| ((!\overline{x_0}\langle 0 \rangle)|x_0(x_0).x_1(x_1).P +_{\rho'_0} (!\overline{x_0}\langle 1 \rangle)|x_0(x_0).x_1(x_1).P) \\
        &\rightarrow_{p_1} \overline{x_1}\langle 0,1,\rho'_{-0} \rangle|(!\overline{x_0}\langle 0 \rangle)|x_1(x_1).P[0/x_0] \\
        &\rightarrow_{p_1} (!\overline{x_0}\langle 0 \rangle)|\overline{x_1}\langle 0,1,\rho'_{-0} \rangle|x_1(x_1).P[0/x_0] \\
        &\rightarrow_{p_1 p_1'} (!\overline{x_0}\langle 0\rangle)|(!\overline{x_1}\langle 0 \rangle)|P[0/x_0, 0/x_1]\\
    \end{split}
\end{equation*}

Here $\rho$ is the density matrix corresponding to the $\ket{00}$
It is important to note here that $p_1=p_1'=1$, since CNOT applied to $\ket{00}$ returns $\ket{00}$. 
\section{State of the Art}\label{sec:related}
This section discusses other work that has focused on lambda calculi for quantum computing. 
One prolific author in this area is Beno\^it Valiron~\cite{benoit_url}. The work that Selinger and Valiron did on quantum lambda calculus~\cite{Selinger_Valiron_2009} can be seen a precursor to Quipper~\cite{quipper_url}. 
Valiron's work in particular~\cite{benoit_thesis} focuses on mathematical machinery that includes category theory and intuitionistic logic and is used to firstly define a lambda calculus for quantum computing, and secondly to mathematically prove certain properties of said lambda calculus.
An approach that closely follows that of Valiron and Selinger is to use von Neumann algebras to describe the denotational semantics of a quantum lambda calculus that only slightly differs from that of Selinger and Valiron~\cite{cho2016vonneumannalgebrasform}.

Another approach preferred by several researchers was to use linear logic to create linear-algebraic lambda calculi ~\cite{DBLP:journals/corr/ArrighiDV13, DBLP:journals/corr/abs-1806-09236, arrighi2005linearalgebraiclambdacalculus, Arrighi2006LinealAL, DBLP:journals/corr/Atzemoglou14, van_Tonder_2004}.
A good and early example of this is van Tonder's lambda calculus~\cite{van_Tonder_2004}. One can find s brief overview of what linear logic is and how it relates to lambda calculus and quantum computing~\cite{vect_lambda_calculus} as well as a more comprehensive treatment of linear logic ~\cite{GIRARD19871}. An interesting example of this approach is one consisting of a type system for linear-algebraic lambda calculus that is especially suited for encoding matrices and vectors~\cite{vect_lambda_calculus}. Linear logic was also used to develop a language for dagger compact categories, called dagger lambda calculus ~\cite{DBLP:journals/corr/Atzemoglou14}. Dagger compact categories are a good candidate for providing an axiomatic framework for quantum protocols and are thus of interest for quantum computing. 

One of the earliest attempts to extend the lambda calculus such that it could express quantum algorithms dates back to 1997~\cite{maymin1997lambdaqcalculusefficientlysimulate, maymin1997extendinglambdacalculusexpress}. The $\lambda^p$-calculus~\cite{maymin1997extendinglambdacalculusexpress} is a $\lambda$-calculus used to express random functions. The $\lambda^q$ calculus is an extension of this and is capable of simulating quantum Turing machines. 

Another approach, derived from linear logic, is to use linear/non-linear (LNL) logic to create a lambda calculus for string diagrams~\cite{DBLP:journals/corr/abs-1804-09822}. String diagrams have applications in a variety of fields, ranging from concurrency theory where they are used to model Petri nets to quantum computing, where they are used to represent quantum circuits. 

Yet another approach that stemmed from linear logic is Quantum FPC~\cite{10.1145/3632855}. Quantum FPC is a combination of FPC~\cite{316083} and the quantum lambda calculus introduced by Selinger and Valiron. 

Another way to construct a lambda calculus for quantum computing is to create a programming language that can manipulate both classical and quantum information~\cite{10.1145/3498687}. The idea is to allow for the use of hybrid classical-quantum algorithms. This is done by using Probabilistic FixPoint Calculus (PFPC) for the classical subsystem and a category of von Neumann algebras to describe quantum effects.

A common strategy in developing quantum lambda calculi is the use of the quantum data - classical control scheme. An example of this is the untyped lambda calculus called Q, which extends by means of a measurement operator to  Q* ~\cite{DALLAGO2011251}. There is work on the quantum control approach as well~\cite{D_az_Caro_2022}. Also in the quantum data - classical control scheme the lambda calculi $\lambda_\rho$ and $\lambda^\circ_\rho$ were defined ~\cite{DBLP:journals/corr/Diaz-Caro17}. In particular $\lambda^\circ_\rho$ makes use of density matrices. \\

A particularly interesting example of a programming language for quantum computing is Qunity~\cite{voichick_2022_7382711}. This is a typed language, with types (and typing rules) pertaining to both expressions and programs. While the language does contain the syntactical notion of abstraction - denoted by $\lambda$ - the semantics for it is a bit different than that of classical lambda calculus. Another thing to note is that the authors also discuss an implementation of Qunity for OpenQASM in their work. While associated with $\lambda$ calculus to an extent, Qunity stands on its own in the space of quantum programming languages.

One of the goals of the Rhyme project is the ability to work with both qbits and p-bits. It is therefore important to also consider models of probabilistic lambda calculus. One approach for this is Probabilistic PCF, described through game semantics~\cite{10.1145/3209108.3209187}. PCF stands for Programmable Computable Functions and is a typed functional language created by Gordon Plotkin. 

Another approach for this is the language $\Lambda^+$, which is an extension of the classical $\lambda$ calculus by adding a probabilistic choice operator.

Yet another approach consists of taking a typed $\lambda$ calculus equipped with operators for sampling from a continuous distribution, recursion and operational semantics described by means of mathematical notions of measure and integration theory. Such models are also called geometry of interaction models for Bayesian programming~\cite{Dal_Lago_Hoshino_2021}.

One more way to approach this problem is through Probabilistic Event Lambda Calculus~\cite{10.1007/978-3-030-45231-5_8}. In this case one takes the classical $\lambda$-calculus with a probabilistic choice operator, such that the operator can be decomposed into a generator, which represents the probabilistic event and a consumer, which acts on the term depending on said event.

Another way to work with probabilistic programming is by using Gradual Probabilistic Lambda Calculus. In this model the probabilistic choice operator can be used to gradually introduce (or remove) static types and probability annotations.

It is worth noting that much work done on quantum lambda calculus comes from the field of mathematics, rather than computer science. 

Th $\pi$-calculus~\cite{MILNER19921,milner_book} is another fundamental tool used for theoretical specification in computer science, but, unlike the $\lambda$-calculus, it is focused on concurrency.

Aside from the original version, the $\pi$-calculus has seen numerous updates and modifications. One such modification was used for the design of secure communication protocols using the Linda shared space model~\cite{10.1016/S0167-6423(02)00090-4}.

Another modification is called broadcast calculus, and is used for the modeling of reconfigurable communicating systems~\cite{925136}.

Yet another extension of the $\pi$-calculus, called the applied $\pi$-calculus was used to model security protocols~\cite{abadi2017appliedpicalculusmobile}. In this calculus values can be obtained from names via predefined functions and equations.

Implementations of the $pi$-calculus have also been created. One of the more well known implementations is called Pict~\cite{6283888}, and it was written in C.

Using the $pi$-calculus in combination with quantum computing has been attempted before~\cite{gay2004communicatingquantumprocesses}.

\section{Calculus}\label{sec:results}
This section presents the core elements of the PiQuant language: the syntax and semantics. Gate operations will be presented for single qubit and multi-qubit systems. The measurement process will also be described, along with the probabilistic choice operator.
The investigation begins with the syntax of the PiQuant language, in Table~\ref{table:piquant_syntax}. 
Precise evaluation semantics is presented in Tables~\ref{tab:semantics1} and~\ref{tab:semantics2}.

\subsection{Core elements}\label{sub:core_elements}

From a theoretical perspective, the syntax of the language is a simple extension of the $\pi$-calculus. 
It contains all standard operators such as process composition ($|$), sending values along a channel, reading a value on a channel, replication, creation of new channels ($\nu$), as well as basic control flow using if statements and boolean expressions.
The syntax is then extended with quantum variables (reading, writing, superposition...).

\begin{table*}[h!]
 	\begin{tabular}{llclp{5cm}} 
 		
 		 Values & $y, y_0, y_1...$ &::= & $x,x_0,x_1...$ & channel names \\ [0.5ex] 
 		 & & $\vert$ &$bexp$ &  boolean expressions \\
 		 Processes & $P,Q,R$ &::=& $x(y).P$ &reads  a value in classical mode, then runs P  \\
 		 & & $\vert$ & $\overline{x}\langle y \rangle.P$ & written classical value \\
 		 & & $\vert$ &  $x[x_1,...,x_n].P$ & state modifier  application, x is a quantum gate name, $x_1$ to $x_n$ are quantum bits \\
 		 & & $\vert$ &  $P\mid Q$ & run P and Q simultaneously \\
 		 & & $\vert$ & $(\nu x)P$ & create a new channel x in P \\
 		 & & $\vert$ & $!P$ & repeatedly spawns copies of P \\
 		 & & $\vert$ & $(\textbf{if} \text{ } bexp \text{ } \textbf{then} \text{ } P \text{ } \textbf{else} \text{ } Q).R$ & if statements \\
 		 & & $\vert$ & $Q\_VAR.P$ & quantum variables \\
 		 & & $\vert$ & $\mathbb{O}$ & terminate the process \\
 		 Quantum variables & $Q\_VAR$ & ::= & $\overline{x}\langle y_1,\alpha \rangle \| \overline{x}\langle y_2,\beta \rangle$ & simple quantum value with weights \\
 		 & & $\vert$ & $\overline{x}\langle y_1, y_2,\rho\rangle$ & simple quantum value with density matrix \\
 		 & & $\vert$ & $\overline{x_0x_1...x_n}\langle y_{01}, y_{02},...,y_{n1}, y_{n2},\rho\rangle$ & multi-qubit state with density matrix \\
 		 Complex variables & $\alpha, \beta, \ldots$& ::= & $v_1+i*v_2$ & $v_1$ and $v_2$ being real values \\
 		 Bit expressions & $bexp$ & ::= & $0,1$ & bit values \\
 		  & & $\vert$ & $NOT \text{ } y$ & negation \\
 		  & & $\vert$ & $y_1 \text{ } AND \text{ } y_2$ & conjunction\\
 		  & & $\vert$ & $y_1 \text{ } OR \text{ } y_2$ &  disjunction
 		
 	\end{tabular}
 	\caption{PiQuant syntax}
 	\label{table:piquant_syntax}
 \end{table*}

    
\begin{table*}[ht]
\begin{center}
\begin{tabular}{c r}
\multicolumn{2}{l}{\fbox{$t\rightarrow t'$}} \\

\\

$\overline{x}\langle y_1, y_2, \rho \rangle | x(y).P \rightarrow ((!\overline{x} \langle y_1 \rangle)|x(y).P ) +_{\rho} ((!\overline{x} \langle y_2 \rangle)|x(y).P)$&(Q-MEASUREMENT)\\
\\
\\
$\frac{p1=tr(|0\rangle\langle 0|\rho)}{((!\overline{x} \langle y_1 \rangle)|x(y).P ) +_{\rho} ((!\overline{x} \langle y_2 \rangle)|x(y).P) \rightarrow_{p1} ((!\overline{x} \langle y_1 \rangle)|x(y).P )}$&(CHOICE-1)\\
\\
\\
$\frac{p1=tr(|0\rangle\langle 0|\rho)}{((!\overline{x} \langle y_1 \rangle)|x(y).P ) +_{\rho} ((!\overline{x} \langle y_2 \rangle)|x(y).P) \rightarrow_{1-p_1} ((!\overline{x} \langle y_2 \rangle)|x(y).P )}$&(CHOICE-2)\\
\\
\\

$\frac{\splitfrac{U_a=|a\rangle\langle a|\otimes \mathbb{1}_{2^n}, \rho_0=trace([1,\ldots,n],\rho), \rho'_{m0}=U_0\rho U_0^\dagger/Tr(U_0\rho U_0^\dagger),}{ \rho'_{m1}=U_1\rho U_1^\dagger/Tr(U_1\rho U_1^\dagger), \rho_{-0}=trace(0, \rho'_{m0}), \rho_{-1}=trace(0, \rho'_{m1})}}{
 \splitfrac{\overline{x_0x_1...x_n}\langle y_{01}, y_{02},...,y_{n1}, y_{n2},\rho\rangle|x_0(y).P \rightarrow}{\splitfrac{
 (\overline{x_1...x_n}\langle y_{11}, y_{12},...,y_{n1}, y_{n2},\rho_{-0}\rangle|  (!\overline{x_0} \langle y_{01} \rangle)| x_0(y).P )}{ +_{\rho_ 0} (\overline{x_1...x_n}\langle y_{11}, y_{12},...,y_{n1}, y_{n2},\rho_{-1}\rangle|(!\overline{x_0} \langle y_{02} \rangle)|x_0(y).P)}}}
 $
&(Q*-MEASUREMENT) 
\\
\\
\end{tabular}
\end{center}

\caption{Semantics of PiQuant - measurement}\label{tab:semantics1}
    \end{table*}


\begin{table*}[ht]
\begin{center}
\begin{tabular}{c r}
\multicolumn{2}{l}{\fbox{$t\rightarrow t'$}} \\

$\frac{0 \leq i \leq n, \text{ }A= \big(\bigotimes_{k=0}^{i-1} \mathbb{1}_2 \big) \otimes \begin{pmatrix}
        0 & 1 \\
        1 & 0
    \end{pmatrix} \otimes \big(\bigotimes_{k=i+1}^{n} \mathbb{1}_2 \big), \text{ } \rho'=A\rho A^\dagger}{\overline{x_0x_1...x_n}\langle y_{01}, y_{02},...y_{n1}, y_{n2}, \rho \rangle | X[x_i].P \rightarrow \overline{x_0x_1...x_n}\langle y_{01}, y_{02},...y_{n1}, y_{n2},  \rho' \rangle |P}$&(X-GATE-APP)
\\
\\

$\frac{0 \leq i \leq n, \text{ }A= \big(\bigotimes_{k=0}^{i-1} \mathbb{1}_2 \big) \otimes \begin{pmatrix}
         0 & -i \\
        i & 0
    \end{pmatrix} \otimes \big(\bigotimes_{k=i+1}^{n} \mathbb{1}_2 \big), \text{ } \rho'=A\rho A^\dagger}{\overline{x_0x_1...x_n}\langle y_{01}, y_{02},...y_{n1}, y_{n2}, \rho \rangle | Y[x_i].P \rightarrow \overline{x_0x_1...x_n}\langle y_{01}, y_{02},...y_{n1}, y_{n2},  \rho' \rangle |P}$&(Y-GATE-APP)
\\
\\

$\frac{0 \leq i \leq n, \text{ }A= \big(\bigotimes_{k=0}^{i-1} \mathbb{1}_2 \big) \otimes \begin{pmatrix}
         1 & 0 \\
        0 & -1
    \end{pmatrix} \otimes \big(\bigotimes_{k=i+1}^{n} \mathbb{1}_2 \big), \text{ } \rho'=A\rho A^\dagger}{\overline{x_0x_1...x_n}\langle y_{01}, y_{02},...y_{n1}, y_{n2}, \rho \rangle | Z[x_i].P \rightarrow \overline{x_0x_1...x_n}\langle y_{01}, y_{02},...y_{n1}, y_{n2},  \rho' \rangle |P}$&(Z-GATE-APP)
\\
\\

$\frac{0 \leq i \leq n, \text{ }A= \big(\bigotimes_{k=0}^{i-1} \mathbb{1}_2 \big) \otimes \begin{pmatrix}
         1 & 0 \\
        0 & e^{2 \pi i /2^k}
    \end{pmatrix} \otimes \big(\bigotimes_{k=i+1}^{n} \mathbb{1}_2 \big), \text{ } \rho'=A\rho A^\dagger}{\overline{x_0x_1...x_n}\langle y_{01}, y_{02},...y_{n1}, y_{n2}, \rho \rangle | R_k[x_i].P \rightarrow \overline{x_0x_1...x_n}\langle y_{01}, y_{02},...y_{n1}, y_{n2},  \rho' \rangle |P}$&($R_k$-GATE-APP)
\\
\\

$\frac{0 \leq i \leq n, \text{ }A= \big(\bigotimes_{k=0}^{i-1} \mathbb{1}_2 \big) \otimes \begin{pmatrix}
         1 & 1 \\
        1 & -1
    \end{pmatrix} \otimes \big(\bigotimes_{k=i+1}^{n} \mathbb{1}_2 \big), \text{ } \rho'=A\rho A^\dagger}{\overline{x_0x_1...x_n}\langle y_{01}, y_{02},...y_{n1}, y_{n2}, \rho \rangle | H[x_i].P \rightarrow \overline{x_0x_1...x_n}\langle y_{01}, y_{02},...y_{n1}, y_{n2},  \rho' \rangle |P}$&(H-GATE-APP)
\\
\\

$\frac{0 \leq i \leq n-1, \text{ }A= \big(\bigotimes_{k=0}^{i-1} \mathbb{1}_2 \big) \otimes \begin{pmatrix}
         1& 0& 0& 0\\
    0& 1& 0& 0\\
    0& 0& 0& 1\\
    0& 0& 1& 0
    \end{pmatrix} \otimes \big(\bigotimes_{k=i+2}^{n} \mathbb{1}_2 \big), \text{ } \rho'=A\rho A^\dagger}{\overline{x_0x_1...x_n}\langle y_{01}, y_{02},...y_{n1}, y_{n2}, \rho \rangle | CNOT[x_i, x_{i+1}].P \rightarrow \overline{x_0x_1...x_n}\langle y_{01}, y_{02},...y_{n1}, y_{n2},  \rho' \rangle |P}$&(CNOT-GATE-APP)
\\
\\

$\frac{0 \leq i \leq n-1, \text{ }A= \big(\bigotimes_{k=0}^{i-1} \mathbb{1}_2 \big) \otimes \begin{pmatrix}
         1& 0& 0& 0\\
    0& 0& 1& 0\\
    0& 1& 0& 0\\
    0& 0& 0& 1
    \end{pmatrix} \otimes \big(\bigotimes_{k=i+2}^{n} \mathbb{1}_2 \big), \text{ } \rho'=A\rho A^\dagger}{\overline{x_0x_1...x_n}\langle y_{01}, y_{02},...y_{n1}, y_{n2}, \rho \rangle | SWAP[x_i, x_{i+1}].P \rightarrow \overline{x_0x_1...x_n}\langle y_{01}, y_{02},...y_{n1}, y_{n2},  \rho' \rangle |P}$&(SWAP-GATE-APP)
\\
\\

$\frac{0 \leq i \leq n-1, \text{ }A= \big(\bigotimes_{k=0}^{i-1} \mathbb{1}_2 \big) \otimes \begin{pmatrix}
         1& 0& 0& 0\\
    0& 1& 0& 0\\
    0& 0& 1& 0\\
    0& 0& 0& e^{2\pi i /2^k}
    \end{pmatrix} \otimes \big(\bigotimes_{k=i+2}^{n} \mathbb{1}_2 \big), \text{ } \rho'=A\rho A^\dagger}{\overline{x_0x_1...x_n}\langle y_{01}, y_{02},...y_{n1}, y_{n2}, \rho \rangle | CR_k[x_i, x_{i+1}].P \rightarrow \overline{x_0x_1...x_n}\langle y_{01}, y_{02},...y_{n1}, y_{n2},  \rho' \rangle |P}$&($CR_k$-GATE-APP)
\\
\\
$\frac{\rho = \rho_1 \otimes \rho_2}{\overline{x_0x_1...x_i}\langle y_{01}, y_{02},...y_{i1}, y_{i2}, \rho_1 \rangle | \overline{x_{i+1}x_{i+2}...x_n}\langle y_{(i+1)1}, y_{(i+2)2},...y_{n1}, y_{n2}, \rho_2 \rangle.P\rightarrow \overline{x_0x_1...x_n}\langle y_{01}, y_{02},...y_{n1}, y_{n2}, \rho \rangle.P} $&(MQ-CONCAT)
\\

\end{tabular}
\end{center}

\caption{Semantics of PiQuant - gates}\label{tab:semantics2}
    \end{table*}

The original $\pi$-calculus already contains a mechanism to encode non-determinism: the parallel composition of programs writing on the same channels.
The natural reduction of a $\pi$-calculus expression where two values are written on the same channel would be that any reader could obtain either one of the two values at random:
\begin{figure*}[ht]
  	\begin{bnf}
  		$P,Q$ : \text{Processes} ::= 
  		| $\overline{x} \langle y \rangle.P$ : write value $y$ on channel $x$, continue with P
  		| $x(y).P$ : read on channel $x$ the value $y$. continue with P
  		| $P \mid Q$ : P and Q run concurrently
  		| $(\nu x)P$ : create new channel $x$, then continue with P
  		| $!P$ : broadcast P
  		| $0$ : terminate the process
  		;; 		
  	\end{bnf}
  	\caption{$\pi$-calculus syntax}
  \end{figure*}

  \begin{table*}[ht]
  	\begin{center}
  		\begin{tabular}{c r}
  			\multicolumn{2}{l}{}\\
                    $\overline{x}\langle z \rangle.P|x(y).P \rightarrow P|Q[z/y]$&(WRITE-READ)\\
  					\\
  					\\
  					$\frac{P \rightarrow Q}{P|R \rightarrow Q|R}$&(RED-COMP)\\
  					\\
  					\\
  					$\frac{P \rightarrow Q}{(\nu x)P \rightarrow (\nu x)Q}$&(RED-NEW)\\
  					\\
  					\\
  					
  					$\frac{\exists x, y \text{ } Q=P[x:=y]}{P \equiv Q}$
  					&(CONG) 
  					\\
  					\\
  					$P|Q \equiv Q|P$&(CONG-PAR)
  					\\
  					\\
  					$(P|Q)|R \equiv P|(Q|R)$&(CONG-ASSOC)
  					\\
  					\\
  					$P|0 \equiv P$&(CONG-PAR-ZERO)
  					\\
  					\\
  					$!P \equiv P|!P$&(CONG-REP)
  					\\
  					\\
  					$\frac{P \equiv P' \text{ } P' \rightarrow Q' \text{ } Q' \equiv Q}{P \rightarrow Q}$&(CONG-RED)
  				\end{tabular}
  			\end{center}
  			
  			\caption{Semantics of the $\pi$-calculus}
  		\end{table*}
\begin{equation*}
    \overline{x}\langle \alpha \rangle | \overline{x}\langle \beta \rangle |(x(x).P) \rightarrow  \overline{x}\langle \beta \rangle |P[\alpha/x]
\end{equation*}

or 
\begin{equation*}
    \overline{x}\langle \alpha \rangle | \overline{x}\langle \beta \rangle |(x(x).P) \rightarrow  \overline{x}\langle \alpha \rangle |P[\beta/x]
\end{equation*}

A subsequent read on the channel $x$ could then read the other value.
This is close to what we need to code quantum variables, but not quite.
What we need is to define the a mechanism that offers the two variables the first time, and then always returns the same read.

The idea of \namesystem~is that quantum variables behave in a way that is close to this mechanism, the difference being that (1) we know from a theoretical standpoint how the choice is made between the two values (superposition of states), and (2) subsequent reads will always read the same value as what was read in the first place (collapse).

To take care of the first aspect, the language defines the superposition operator ($||$), which ties the possible values taken by a quantum variable together, making sure that they cannot be read independently.
The second aspect is taken care of by the Q-MEASUREMENT reduction rule defined in Table~\ref{tab:semantics2}.

In this language, a quantum variable $x$ in a pure state that can take either value $y_1$ or $y_2$ with weights $\alpha$ and $\beta$ is defined as:
\begin{equation*}
    \overline{x}\langle y_1, \alpha \rangle || \overline{x}\langle y_2, \beta \rangle
\end{equation*}

If we add a program $x(x).P$ reading it, the resulting program can be expressed as: 
\begin{equation*}
    \overline{x}\langle y_1, \alpha \rangle || \overline{x}\langle y_2, \beta \rangle |x(x).P
\end{equation*}

One of the  contributions of the present work is to specify how the choice is made using quantum theory.
The way to do it is to prevent readers from reading these values independently.
It prevents reducing from one channel to the other and also allows modifying $\alpha$ and $\beta$ at the same time. $y_1$ and $y_2$ are values in $\{0,1\}$ and $\alpha$ and $\beta$
are normalized complex values.

For convenience, a simple qubit can also be expressed as:
\begin{equation*}
    \overline{x}\langle y_1, y_2,\begin{pmatrix}
        \alpha \alpha^* & \alpha \beta^* \\
        \beta \alpha^* & \beta \beta^*
    \end{pmatrix}\rangle
\end{equation*}

This is equivalent to: 
\begin{equation*}
    \overline{x}\langle y_1, \alpha \rangle || \overline{x}\langle y_2, \beta \rangle
\end{equation*}

\subsection{Applying gates to quantum variables}\label{sub:app_gates}

The application of quantum gates is then a simple modification of the density matrix. For example, the X gate is applied as follows:
\begin{equation*}
    \overline{x}\langle y_1, y_2, \rho \rangle | X[x].P \rightarrow \overline{x}\langle y_1, y_2, X\rho X^\dagger \rangle |P
\end{equation*}

The Y gate is applied as follows:
\begin{equation*}
    \overline{x}\langle y_1, y_2, \rho \rangle | Y[x].P \rightarrow \overline{x}\langle y_1, y_2, Y\rho Y^\dagger \rangle | P
\end{equation*}

The Z gate is defined as:
\begin{equation*}
    \overline{x}\langle y_1, y_2, \rho \rangle | Z[x].P \rightarrow \overline{x}\langle y_1, y_2, Z\rho Z^\dagger \rangle | P
\end{equation*}

The measurement is defined as:
\begin{equation*}
    \overline{x}\langle y_1, y_2, \rho \rangle | x(y).P \rightarrow ((!x \langle y_1 \rangle)|x(y).P ) +_{\rho} ((!\overline{x} \langle y_2 \rangle)|x(y).P)
\end{equation*}

The Hadamard gate is applied in the following way:
\begin{equation*}
\overline{x}\langle y_1, y_2, \rho \rangle|H[x].P \rightarrow \overline{x}\langle y_1, y_2, H\rho H^\dagger \rangle|P
\end{equation*}

The $R_k$ gate is applied in the following manner:
\begin{equation*}
    \overline{x_0x_1...x_n}\langle y_{01}, y_{02},...y_{n1}, y_{n2}, \rho \rangle | R_k[x_i].P \rightarrow \overline{x_0x_1...x_n}\langle y_{01}, y_{02},...y_{n1}, y_{n2},  \rho' \rangle |P
\end{equation*}

A SWAP operation is performed in the following manner:
\begin{equation*}
    \overline{x_0x_1...x_n}\langle y_{01}, y_{02},...y_{n1}, y_{n2}, \rho \rangle | SWAP[x_i, x_{i+1}].P
\end{equation*}

A multi-qubit state, whether entangled or not is represented in the following way:
\begin{equation*}
    \overline{x_0x_1...x_n}\langle y_{01}, y_{02},...y_{n1}, y_{n2},\rho\rangle|x_0(y).P
\end{equation*}
The $y_{ij}$ are labels. They describe the choices that each qubit in the multi-qubit state represents. They are the way to interpret qubits and multi-qubits in the $\pi$ calculus formalism

The CNOT gate is applied in the following way:
\begin{equation*}
CNOT|\overline{x_0 x_1}\langle y_{01}, y_{02}, y_{11}, y_{12}, \rho \rangle | x(y).P \rightarrow \overline{x_0 x_1}\langle y_{01}, y_{02}, y_{11}, y_{12}, CNOT \rho CNOT^\dagger \rangle | x(y).P
\end{equation*}

The controlled $R_k$, denoted by $CR_k$ is applied in the following manner:
\begin{equation*}
    \overline{x_0x_1...x_n}\langle y_{01}, y_{02},...y_{n1}, y_{n2}, \rho \rangle | CR_k[x_i, x_{i+1}].P \rightarrow \overline{x_0x_1...x_n}\langle y_{01}, y_{02},...y_{n1}, y_{n2},  \rho' \rangle |P
\end{equation*}
One thing to note about the $CNOT$ and $CR_k$ gates is that the order of the arguments matters - the first argument ($x_i$ in the example above) represents the control qubit.

For the gate applications the following known rule was applied, for the evolution of the density matrix:
\begin{equation*}
    \rho'=U\rho U^\dagger
\end{equation*}
Where $U$ is a unitary operator, possibly formed through the tensor product of several unitary operators in lower dimension.

Different multi-qubit states are handled in the following way:

\begin{equation*}
    \frac{\rho = \rho_1 \otimes \rho_2}{\overline{x_0x_1...x_i}\langle y_{01}, y_{02},...y_{i1}, y_{i2}, \rho_1 \rangle | \overline{x_{i+1}x_{i+2}...x_n}\langle y_{(i+1)1}, y_{(i+2)2},...y_{n1}, y_{n2}, \rho_2 \rangle.P\rightarrow \overline{x_0x_1...x_n}\langle y_{01}, y_{02},...y_{n1}, y_{n2}, \rho \rangle.P}
\end{equation*}
    
Separate multi-qubit states are merged together. 
In this way, applying multi-qubit gates is done in the usual way, with no need to take into account which register each qubit belongs to.

When performing measurements, the $+_{\rho}$ operator is a probabilistic choice operator is where the complexity of the quantum entanglement happens. 
The following section describes precisely the $+_{\rho}$ operator.

\subsection{The probabilistic choice operator $+_{\rho}$}\label{square_rho}
The probabilistic choice operator used in the semantic rules pertaining to measurement functions in the following way: the starting point is measurement of the form $M_1=|0\rangle \langle 0|$. 
Then the value $0$ will be obtained with probability $tr(M_1\rho)$, while value $1$ will be obtained with probability $tr(M_2\rho)$, where $M_2=|1\rangle \langle 1|$. $\rho$ is the system's density matrix at the moment the measurement is performed - i.e. gate operations up to the moment the measurement is performed are taken into account. 
The operator $\rightarrow_{p_1}$ used in rule (CHOICE-1) means that the evaluation is done non-deterministically, with probability $p_1$. 
Similarly $\rightarrow_{p_2}$ means that the evaluation in rule (CHOICE-2) is done non-deterministically, with probability $p_2$. For the more generalized rule (Q*-MEASUREMENT), the process is very similar, except the probabilities are determined using matrices obtained by taking partial traces of the main density matrix. More specifically, the matrix $\rho_{-0}$ is obtained by taking partial trace by the subsystem consisting of the first qubit of the density matrix $\rho$. 
The matrix $\rho_0$ is obtained by taking a partial trace of $\rho$, but this time by the subsystem consisting of all other qubits. 
An important assumption is that, in order to obtain a measurement of a particular qubit's state in a multi-qubit system, wires can be permuted in the quantum circuit. Such a permutation would entail a rearrangement of the gates as well. The key notions here are those of graph isomorphism and graph automorphism. 
Two graphs are isomorphic if there is a bijection $f$ between their respective sets of vertices that also preserves adjacency. The quantum circuit is regarded as a graph where the initial qubit states and the gates are vertices, and the wires in the quantum circuit are edges. 
The circuit obtained by permuting wires is isomorphic to the original circuit, in the sense of graph isomorphisms.

\subsection{Restrictions on the density and measurement matrices}

Note that the density and measurement matrices are not random arbitraty matrices. 
Constraints are passed from the definitions of quantum theory.
We list them below:

\begin{itemize}
    \item $tr \rho = 1$, for density matrices of any dimension
    \item $\rho \geq 0$, meaning the density matrices need to be positive semi-definite
    \item $M_m=M_m^\dagger=M_m^2$, meaning the measurement matrices are Hermitian. We assume only single qubit measurements, i.e. the measurement matrices are projectors of dimension $2 \times 2$
\end{itemize}

\section{Examples of Programs}\label{sec:ex_prog}
In quantum computing, programs are typically represented as quantum "circuits", i.e. diagrams that show the transformations that the qubits undergo. This section will showcase some qubit transformations both in diagram and PiQuant form.\\

\subsection{Measuring in a different basis - Hadamard}
Our semantics rules assume measurements are always done in the computational basis. Measuring in a different basis would simply entail applying the correct gate(s) on the qubit to be measured prior to measuring. For example, if a measurement in the Hadamard basis is desired, one simply needs to apply the Hadamard gate on the qubit to be measured. The circuit for such a measurement would be:
\[\Qcircuit @C=1 em @R=1 em {
  \lstick{\ket{x_0}}  & \gate{H} & \meter & \cw
}\]

Written in PiQuant this would be:
\begin{equation}
\begin{split}
    &\overline{x_0}\langle y_{01}, y_{02}, \rho \rangle|H[x_0].x_0(x_0).P\\
    &\rightarrow \overline{x_0}\langle y_{01}, y_{02}, \rho' \rangle|x_0(x_0).P\\
    &\rightarrow (!\overline{x_0}\langle 0 \rangle).P[0/x_0] +_{\rho'} (!\overline{x_0}\langle 1 \rangle).P[1/x_0]\\
    &\rightarrow_{p_1} (!\overline{x_0}\langle 0 \rangle).P[0/x_0]
\end{split}
\end{equation}

where:
\begin{equation*}
    \rho' = H \rho H^\dagger
\end{equation*}

\subsection{H gate}\label{H_ex}
Applying the Hadamard gate can be represented as:
\[\Qcircuit @C=1 em @R=1 em {
  \lstick{q_0 = \ket{0}}  & \ustick{|} \qw & \gate{H} & \ustick{.}\qw & \meter & \ustick{.} \cw & \cw 
}\]
Below is this program written in PiQuant, with the intermediate evaluation steps:
\begin{equation*}
    \begin{split}
        &\overline{x_0}\langle 0,1, \begin{pmatrix}
            1 &0 \\
            0 & 0
        \end{pmatrix} \rangle| H[x_0].x_0(x_0).P \\
        & \rightarrow \overline{x_0}\langle 0,1, H \rho H^\dagger \rangle|x_0(x_0).P \\
        & \rightarrow (!\overline{x_0}\langle  0 \rangle)|x_0(x_0).P +_{\rho'} (!\overline{x_0}\langle  1 \rangle)|x_0(x_0).P\\
        &\rightarrow_{p_1} (!\overline{x_0}\langle 0 \rangle)|P
    \end{split}
\end{equation*}

where :
\begin{equation*}
    \rho = \begin{pmatrix}
            1 &0 \\
            0 & 0
        \end{pmatrix}; \text{ } \rho' = H\rho H^\dagger
\end{equation*}
The following $\pi$-calculus evaluation rule was also used:
\begin{equation*}
    (!\overline{x_0}\langle 0 \rangle)|x_0(x_0).P \rightarrow (!\overline{x_0}\langle 0 \rangle)|P
\end{equation*}
\subsection{CNOT}\label{cnot_ex}
Applying the CNOT gate can be represented as:
\[\Qcircuit @C=1 em @R=1 em {
 \lstick{q_0=\ket{1}} & \ustick{|} \qw& \ctrl{1} &  \ustick{.} \qw& \meter & \ustick{.} \cw&\cw\\
 \lstick{q_1=\ket{0}} & \ustick{|} \qw& \targ  &\ustick{.} \qw& \meter & \ustick{.} \cw&\cw
}\]

Written in PiQuant this becomes:
\begin{equation*}
    \begin{split}
        & \overline{x_0 x_1}\langle0,1,0,1, \begin{pmatrix}
            0 & 0 & 0 & 0 \\
            0 & 0 & 0 & 0 \\
            0 & 0 & 1 & 0 \\
            0 & 0 & 0 & 0 
        \end{pmatrix} \rangle| CNOT[x_0,x_1].x_0(x_0).x_1(x_1).P \\
        & \rightarrow \overline{x_0 x_1} \langle 0,1,0,1, \rho' \rangle|x_0(x_0).x_1(x_1).P\\
        & \rightarrow \overline{x_1}\langle 0,1, \rho'_{-0}\rangle| ((!\overline{x_0}\langle 0 \rangle)|x_0(x_0).x_1(x_1).P +_{\rho'_0} (!\overline{x_0}\langle 1 \rangle)|x_0(x_0).x_1(x_1).P) \\
        &\rightarrow_{p_2} \overline{x_1}\langle 0,1,\rho'_{-0} \rangle|(!\overline{x_0}\langle 1 \rangle)|x_1(x_1).P[1/x_0] \\
        &\rightarrow_{p_2} (!\overline{x_0}\langle 1 \rangle)|\overline{x_1}\langle 0,1,\rho'_{-0} \rangle|x_1(x_1).P[1/x_0] \\
        &\rightarrow_{p_2 p_2'} (!\overline{x_0}\langle 1\rangle)|(!\overline{x_1}\langle 1 \rangle)|P[1/x_0, 1/x_1]\\
    \end{split}
\end{equation*}
It is important to note here that $p_2=p_2'=1$.  
\subsection{SWAP}\label{swap_ex}
Applying the SWAP gate can be represented as:

\[\Qcircuit @C=1 em @R=1 em {
 \lstick{q_0=\ket{0}} & \ustick{|} \qw& \qswap &  \ustick{.} \qw& \meter & \ustick{.} \cw&\cw\\
 \lstick{q_1=\ket{1}} & \ustick{|} \qw& \qswap \qwx &\ustick{.} \qw& \meter & \ustick{.} \cw&\cw
}\]

Written in PiQuant this is:
\begin{equation*}
    \begin{split}
        & \overline{x_0 x_1}\langle0,1,0,1, \begin{pmatrix}
            0 & 0 & 0 & 0 \\
            0 & 1 & 0 & 0 \\
            0 & 0 & 0 & 0 \\
            0 & 0 & 0 & 0 
        \end{pmatrix} \rangle| SWAP[x_0,x_1].x_0(x_0).x_1(x_1).P \\
        & \rightarrow \overline{x_0 x_1} \langle 0,1,0,1, \rho' \rangle|x_0(x_0).x_1(x_1).P\\
        & \rightarrow \overline{x_1}\langle 0,1, \rho'_{-0}\rangle| ((!\overline{x_0}\langle 0 \rangle)|x_1(x_1).P[0/x_0] +_{\rho'_0} (!\overline{x_0}\langle 1 \rangle)|x_1(x_1).P[1/x_0]) \\
        &\rightarrow_{p_2} \overline{x_1}\langle 0,1,\rho'_{-0} \rangle|(!\overline{x_0}\langle 1 \rangle)|x_1(x_1).P[1/x_0] \\
        &\rightarrow_{p_2} (!\overline{x_0}\langle 1 \rangle)|\overline{x_1}\langle 0,1,\rho'_{-0} \rangle|x_1(x_1).P[1/x_0] \\
        &\rightarrow_{p_2 p_1'} (!\overline{x_0}\langle 1\rangle)|(!\overline{x_1}\langle 0 \rangle)|P[1/x_0, 0/x_1]\\
    \end{split}
\end{equation*}

The choices in the example above correspond to the fact that $p_2=p_1'=1$.

\subsection{Bell state "unentangle"}\label{bell_unentangle_ex}
Below is a circuit that maps Bell states to 2-qubit unentangled states in the computational basis. The example provided shows how the $\ket{\phi_+}$ state is mapped to the $\ket{00}$ state. This is an example of how Bell state measurements are done in practice - by using a circuit that maps the entangled states to unentangled ones in the computational basis.
\[\Qcircuit @C=1 em @R=1 em {
 \lstick{} & \ustick{|} \qw& \ctrl{1} &  \ustick{.} \qw& \gate{H} &  \ustick{.} \qw & \meter & \ustick{.} \cw&\cw\\
 \lstick{} & \ustick{|} \qw& \targ  &\ustick{.} \qw& \qw& \qw &\meter & \ustick{.} \cw&\cw
 \inputgroupv{1}{2}{.8em}{.8em}{\ket{\phi_+}}\\
}\]

Written in PiQuant, this is:
\begin{equation*}
    \begin{split}
        & \overline{x_0 x_1}\langle 0,1,0,1,\frac{1}{2} \begin{pmatrix}
            1 & 0 & 0 & 1 \\
            0 & 0 & 0 & 0 \\
            0 & 0 & 0 & 0 \\
            1 & 0 & 0 & 1 
        \end{pmatrix} \rangle| CNOT[x_0,x_1].H[x_0].x_0(x_0).x_1(x_1).P \\
        & \rightarrow \overline{x_0 x_1} \langle 0,1,0,1, \rho' \rangle|H[x_0].x_0(x_0).x_1(x_1).P\\
        & \rightarrow \overline{x_0 x_1} \langle 0,1,0,1, \rho'' \rangle|x_0(x_0).x_1(x_1).P\\
        & \rightarrow \overline{x_1}\langle 0,1, \rho''_{-0}\rangle| ((!\overline{x_0}\langle 0 \rangle)| x_1(x_1).P[0/x_0] +_{\rho''_0} (!\overline{x_0}\langle 1 \rangle)|x_1(x_1).P[1/x_0]) \\
        &\rightarrow_{p_1} \overline{x_1}\langle 0,1,\rho''_{-0} \rangle|(!\overline{x_0}\langle 0 \rangle)|x_1(x_1).P[0/x_0] \\
        &\rightarrow_{p_1} (!\overline{x_0}\langle 0 \rangle)|\overline{x_1}\langle 0,1,\rho''_{-0} \rangle|x_1(x_1).P[1/x_0] \\
        &\rightarrow_{p_1 p_1'} (!\overline{x_0}\langle 0\rangle)|(!\overline{x_1}\langle 0 \rangle)|P[0/x_0, 0/x_1]\\
    \end{split}
\end{equation*}
The choices in this example correspond to the fact that $p_1=p_1'=1$.
The matrices used are defined as follows:
\begin{equation*}
    \rho=\frac{1}{2} \begin{pmatrix}
        1 & 0 & 0 & 1 \\
            0 & 0 & 0 & 0 \\
            0 & 0 & 0 & 0 \\
            1 & 0 & 0 & 1 
    \end{pmatrix}
\end{equation*}

\begin{equation*}
    \rho'=CNOT \rho CNOT^\dagger
\end{equation*}

\begin{equation*}
    U=H\otimes \mathbb{1},\rho''=U \rho U^\dagger
\end{equation*}

\subsection{Z gate}\label{z_ex}
Applying $\sigma_z$ to a qubit $q_0$ can be represented as:

\[\Qcircuit @C=1 em @R=1 em {
  \lstick{\ket{x_0}}  & \gate{Z} & \qwa 
}\]

Written in PiQuant the program would be:
\begin{equation}
    \overline{x_0}\langle y_{01}, y_{02}, \rho \rangle|Z[x_0].P
\end{equation}

where $Z$ corresponds to the $\sigma_z$ matrix.\\

\subsection{CNOT, Hadamard on the second qubit}
Another circuit example is:
\[\Qcircuit @C=1 em @R=1 em {
 \lstick{\ket{x_0}} & \ctrl{1} &  \qw & \qwa\\
 \lstick{\ket{x_1}} & \targ  & \gate{H} & \qwa
}\]
where the first transformation is the 2-qubit CNOT gate and the second is the single qubit Hadamard gate.
This program can be written as:
\begin{equation}
    \overline{x_0 x_1}\langle y_{01}, y_{02},y_{11}, y_{12} \rho \rangle|CNOT[x_0, x_1].H[x_1].P
\end{equation}

\subsection{Quantum Fourier Transform}
The next example program is the quantum Fourier transform:

\[\Qcircuit @C=1 em @R=1 em {
 \lstick{\ket{x_0}} & \gate{H} & \gate{R_2} & \gate{R_3} & \qw & \qw & \qw &\qwa \\
 \lstick{\ket{x_1}} & \qw & \ctrl{-1} & \qw & \gate{H} & \gate{R_2}& \qw &\qwa \\
 \lstick{\ket{x_2}} & \qw & \qw & \ctrl{-2} & \qw & \ctrl{-1}& \qw & \qwa
}\]
This program can be written in the following way:
\begin{equation}
\begin{split}
    \overline{x_0 x_1 x_2}\langle y_{01}, y_{02},y_{11}, y_{12}, y_{21}, y_{22}, \rho \rangle|H[x_0] & \\.CR_2[x_1, x_0].CR_3[x_2, x_0].H[x_1].CR_2[x_2, x_1].P
 \end{split}   
\end{equation}

\subsection{Quantum error correction}
This subsection treats a classical example pertaining to quantum error correction, the encoding for the 3 qubit bit flip code. The circtuit for this encoding is illustrated in the circuit diagram below:
\[\Qcircuit @C=1 em @R=1 em {
 \lstick{\ket{\psi}} & \qw & \ctrl{1} &  \ctrl{2} & \qw & \qwa \\
 \lstick{\ket{0}} & \qw & \targ & \qw &  \qw &\qwa \\
 \lstick{\ket{0}} & \qw & \qw & \targ & \qw & \qwa
}\]

Written in PiQuant this would be:
\begin{equation*}
    \overline{x_0 x_1 x_2}\langle y_{01}, y_{02},y_{11}, y_{12}, y_{21}, y_{22}, \rho \rangle|CNOT[x_0,x_1]|CNOT[x_0,x_2].P
\end{equation*}

\subsection{Quantum communication: BB84 encryption}
\[\Qcircuit @C=1 em @R=1 em {
 \lstick{d} & \cctrl{1} \\
 \lstick{b} & \cctrl{1} \\
 \lstick{\ket{0}} & \qw & \ustick{00} \qw & \qw & \qwa\\
 \lstick{} & & \ustick{01} \qw & \qw & \gate{H} & \qwa\\
 \lstick{} & & \ustick{10} \qw & \qw & \gate{X} & \qwa\\
 \lstick{} & \push{} \cwx[-3] & \ustick{11} \qw & \qw & \gate{X} & \gate{H} & \qwa\\
}\]

This subsection focuses on a particular segment of the BB84 protocol~\cite{nielsen_chuang}, namely the way in which Alice chooses which qubits to send to Bob. Consider two classical bits, $d$ a data bit and $b$ a bit that indicates the basis for the qubit to be sent. If $b=0$, the qubit is sent in the $\{\ket{0}, \ket{1}\}$ basis, otherwise the qubit is sent in the $\{\ket{+}, \ket{-}\}$ basis. If $b=0$, then $d=0$ corresponds to $\ket{0}$ and $d=1$ corresponds to $\ket{1}$. If $b=1$, then $d=0$ corresponds to $\ket{+}$ and $d=1$ corresponds to $\ket{-}$.
The circuit above shows this correspondence.\\
In BB84, once Alice has her data bit and encoding established, she sends her qubit to Bob, and he measures it arbitrarily in either the computational or Hadamard basis. Bob's part of the circuit would be represented in the followig way:

\[\Qcircuit @C=1 em @R=1 em {
 \lstick{rnd} & \cctrl{1} \\
 \lstick{A} & \qw & \ustick{0} \qw & \qw & \meter\\
 \lstick{} &\push{} \cwx[-1] & \ustick{1} \qw & \qw & \gate{H} & \meter\\
}\]

Here $A$ represents the qubit sent by Alice, and $rnd$ is a random number that takes values 0 or 1. If $rnd = 0$, then Bob measures in the computational basis, otherwise he measures in the Hadamard basis.

Putting it all together we get the following circuit:
\[\Qcircuit @C=1 em @R=1 em {
 \lstick{d} & \cctrl{1} \\
 \lstick{b} & \cctrl{1} \\
 \lstick{rnd} & \cctrl{1} \\
 \lstick{\ket{0}} & \qw & \ustick{000} \qw & \qw & \meter\\
 \lstick{} & & \ustick{001} \qw & \qw & \gate{H} & \meter\\
 \lstick{} & & \ustick{010} \qw & \qw & \gate{H} & \meter\\
 \lstick{} & & \ustick{011} \qw & \qw & \qw & \meter\\
 \lstick{} & & \ustick{100} \qw & \qw & \gate{X} & \meter\\
 \lstick{} & & \ustick{101} \qw & \qw & \gate{X} & \gate{H} & \meter\\
 \lstick{} &  & \ustick{110} \qw & \qw & \gate{X} & \gate{H} & \meter\\
 \lstick{} & \push{} \cwx[-7] & \ustick{111} \qw & \qw & \gate{X} & \qw & \meter\\
}\]

Written in PiQuant, this would be:
\begin{equation*}
\begin{split}
    &(\textbf{if not }d\textbf{ and not }b \textbf{ and not }rnd \textbf{ then }\\
    &\overline{x_0}\langle y_0, y_1, \rho \rangle|x_0(x_0) \\
    &\textbf{else }\\
    &\textbf{if not }d\textbf{ and not }b \textbf{ and }rnd \textbf{ then }\\
    &\overline{x_0}\langle y_0, y_1, \rho \rangle|H[x_0].x_0(x_0) \\
    &\textbf{else }\\
    &\textbf{if not }d\textbf{ and }b\textbf{ and not }rnd\textbf{ then } \\
    &\overline{x_0}\langle y_0, y_1, \rho \rangle|H[x_0].x_0(x_0) \\
    &\textbf{else } \\
    &\textbf{if not }d\textbf{ and }b\textbf{ and }rnd\textbf{ then } \\
    &\overline{x_0}\langle y_0, y_1, \rho \rangle|x_0(x_0) \\
    &\textbf{else } \\
    &\textbf{if }d\textbf{ and not }b\textbf{ and not }rnd \textbf{ then } \\
    &\overline{x_0}\langle y_0, y_1, \rho \rangle|X[x_0].x_0(x_0) \\
    &\textbf{else } \\
    &\textbf{if }d\textbf{ and not }b\textbf{ and }rnd \textbf{ then } \\
    &\overline{x_0}\langle y_0, y_1, \rho \rangle|X[x_0].H[x_0].x_0(x_0) \\
    &\textbf{else } \\
    &\textbf{if }d\textbf{ and }b\textbf{ and not }rnd\textbf{ then } \\
    &\overline{x_0}\langle y_0, y_1, \rho \rangle|X[x_0].H[x_0].x_0(x_0)\\
    &\textbf{else }\\
    &\textbf{if }d\textbf{ and }b\textbf{ and }rnd\textbf{ then } \\
    &\overline{x_0}\langle y_0, y_1, \rho \rangle|X[x_0].x_0(x_0)\\
    &).P
\end{split}
\end{equation*}

\section{Implementation}\label{sec:implementation}
The PiQuant interpreter is available for download.\footnote{\url{https://github.com/CatalinT25/PiQuant}}
It consists of an interpreter written in OCaml. 
The implementation is relatively compact and the total number of lines of code is 1570 grouped in 54 functions. 
This includes the reduction logic, parsing and interpreter code. 
While PiQuant does not formally have types and typing rules, the implementation reuses some OCaml types. 
Notably this implementation uses the Owl module for matrix operations because of its efficiency. 

This implementation was created in Ubuntu Linux and designed to work as an interpreter running in the terminal. 
The way it works is the following: the user inputs an expression in PiQuant, the interpreter reduces that expression until there are no more reducible steps. 
There is one exception to this and that is broadcasting. 
One could in principle apply the broadcasting rule infinitely many times. This works in a theoretical setting, but not in a practical one. Therefore, when one has a broadcast of the type:
\begin{equation*}
	!a1<0>.P
\end{equation*}
where  $P$ contains no reads, i.e. no terms like $a1(x)$, then the reduction stops, and the broadcast is simply displayed as shown above. It is worth noting here that broadcasts are, at least in the current version of PiQuant, only done on classical bits. 
This is done in order to avoid complications related to the no-cloning theorem. 
What follows in this section is 
examples of input-output sequences will be shown.

\subsection{Examples}
Some examples of input and output in the PiQuant interpreter are shown in what follows:

\begin{figure}[H]
	\includegraphics[scale=0.3]{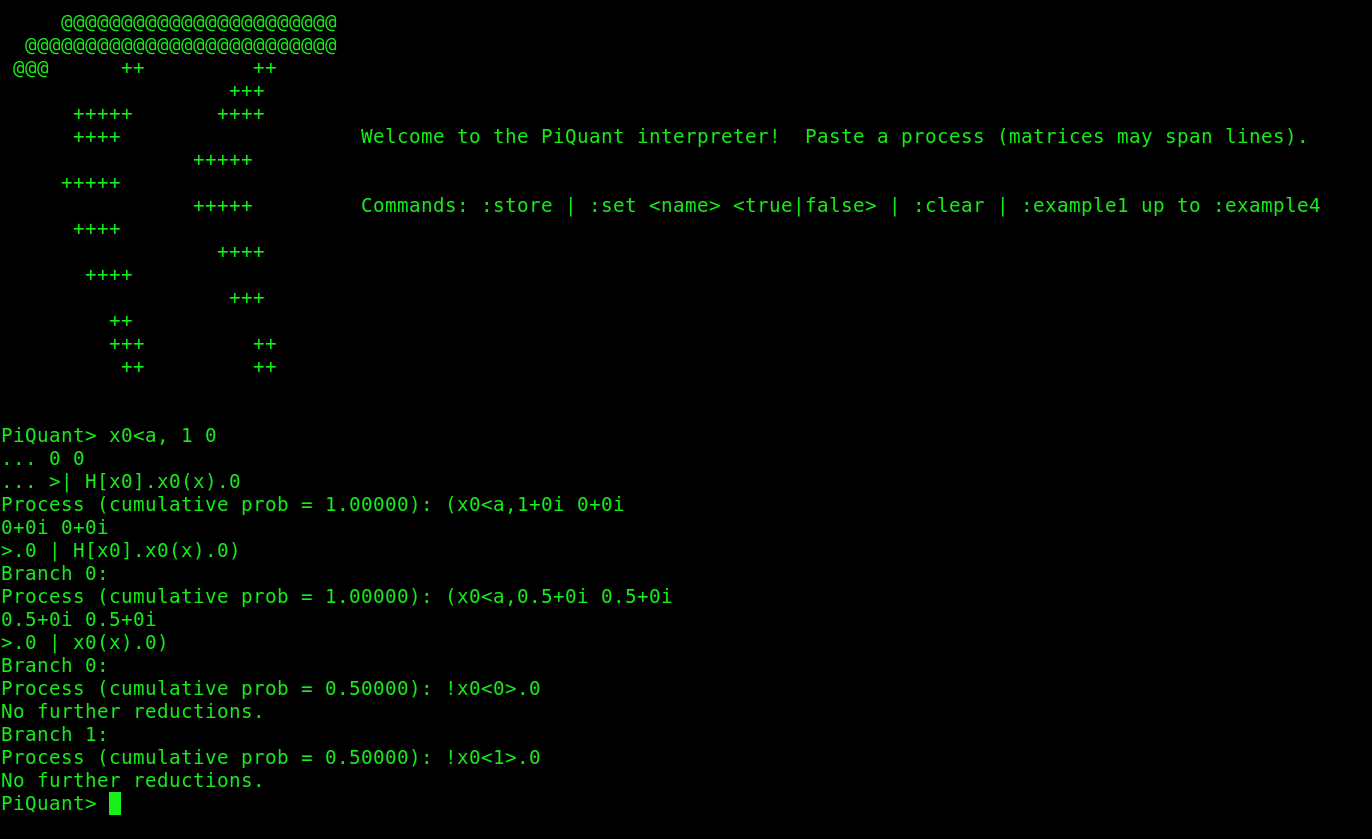}
\end{figure}

This example corresponds to \ref{H_ex}. The gate H is applied to the state $\ket{0}$, and then a measurement is performed. The result obtained matches the theoretical result presented in \ref{H_ex}

\begin{figure}[H]
	\includegraphics[scale=0.3]{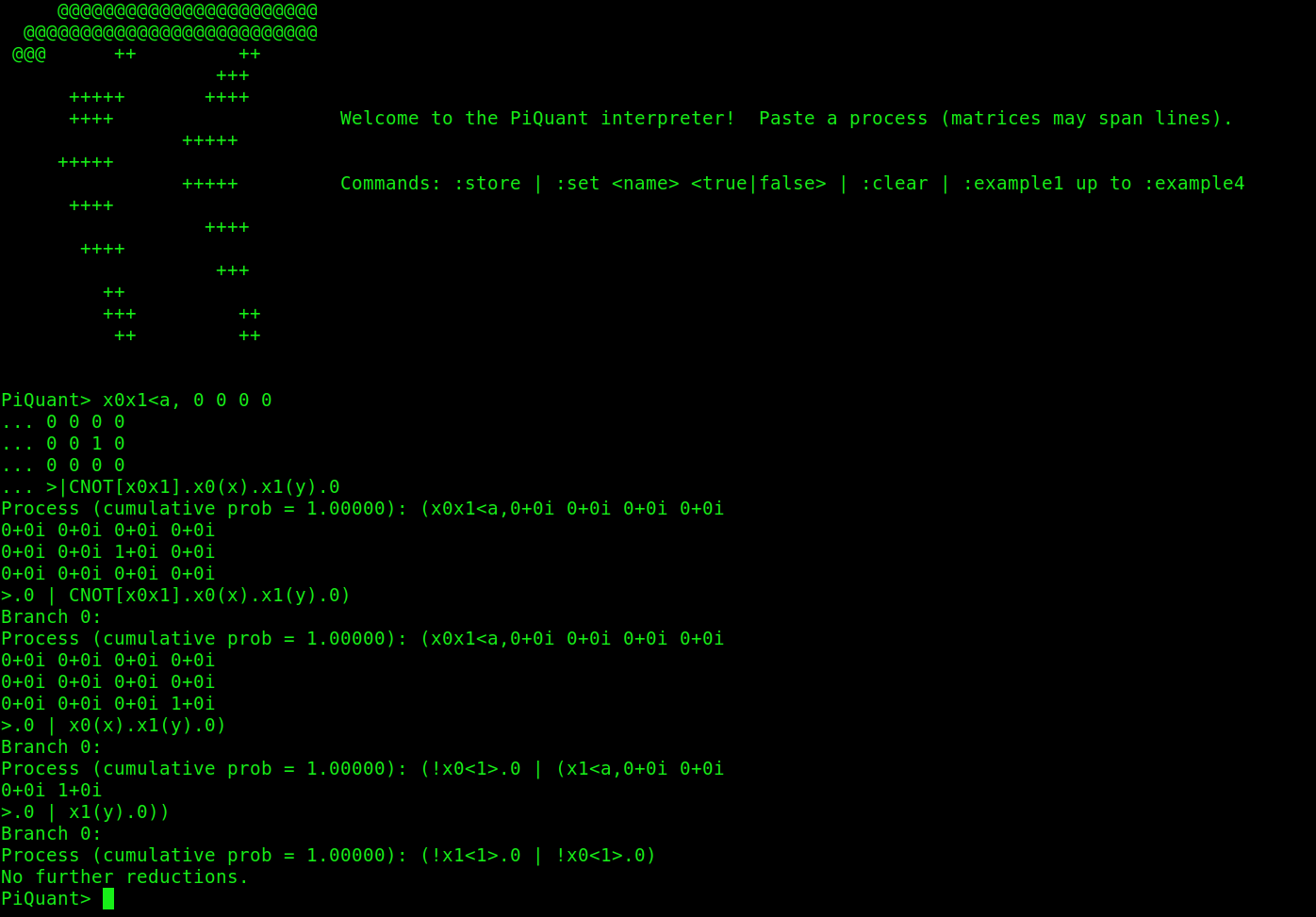}
\end{figure}

This example corresponds to \ref{cnot_ex}. The CNOT gate is applied to the state:
\begin{equation*}
	\ket{10}  = \begin{pmatrix}
		0 \\ 
		0 \\ 
		1 \\
		0
	\end{pmatrix}
\end{equation*}

whose corresponding density matrix is:
\begin{equation*}
	\rho = \begin{pmatrix}
		0 &0 &0 &0 \\
		0 &0 &0 &0 \\
		0 &0 &1 &0 \\
		0 &0 &0 &0 
	\end{pmatrix}
\end{equation*}

Subsequently, measurements are performed on the two qubits. The result obtained is in line with the theoretical result obtained in \ref{cnot_ex}.

\begin{figure}[H]
	\includegraphics[scale=0.3]{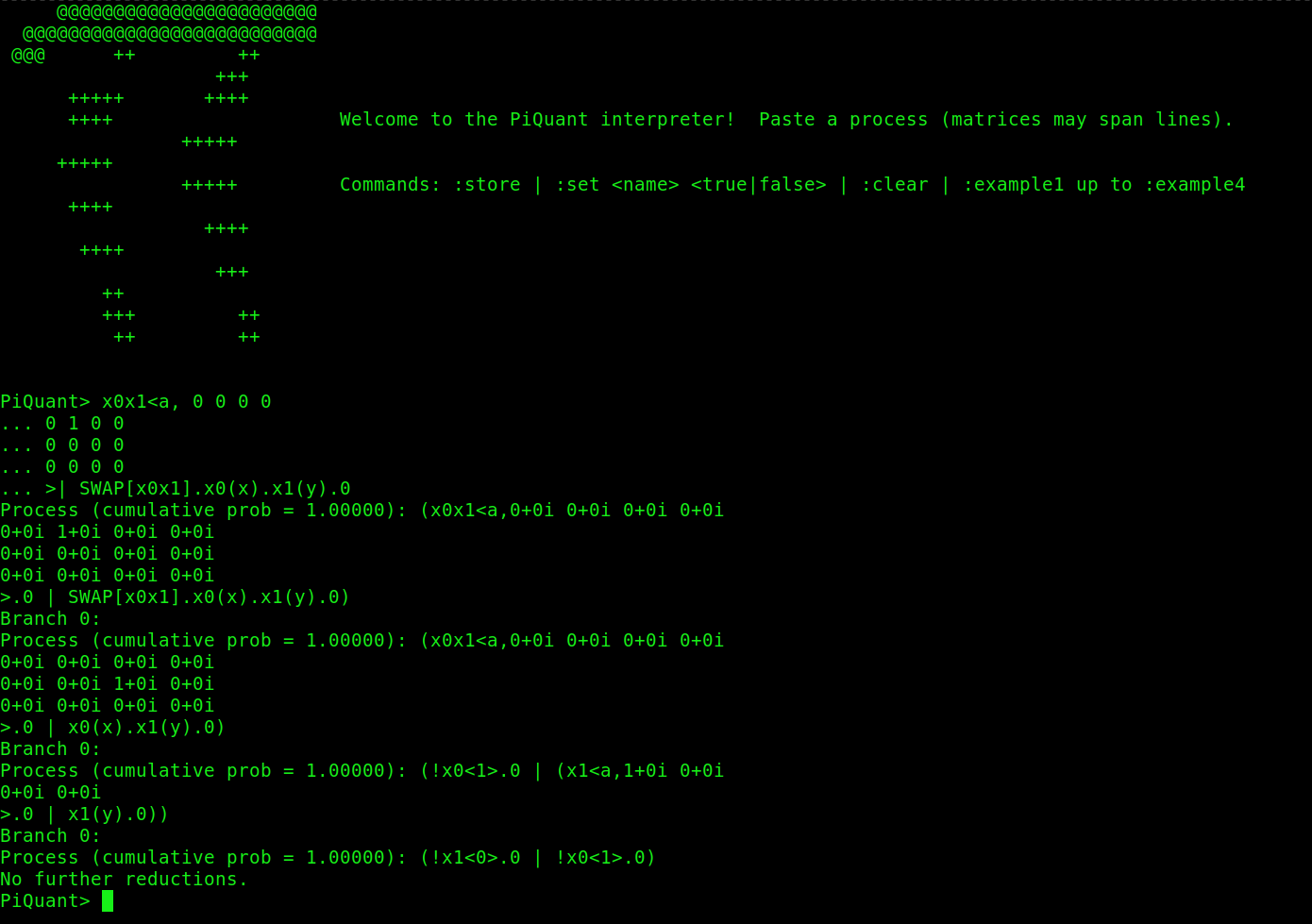}
\end{figure}

This example corresponds to \ref{swap_ex}. The SWAP gate is applied to the state:
\begin{equation*}
	\ket{01}= \begin{pmatrix}
		0 \\
		1\\
		0 \\
		0
	\end{pmatrix}
\end{equation*}
whose density matrix is:
\begin{equation*}
	\rho = \begin{pmatrix}
		0 &0 &0 &0 \\
		0 &1 &0 &0 \\
		0 &0 &0 &0 \\
		0 &0 &0 &0 
	\end{pmatrix}
\end{equation*}
Measurements on the two qubits are then performed. The result obtained is in line with the theoretical result obtained in \ref{swap_ex}.

\begin{figure}[H]
	\includegraphics[scale=0.3]{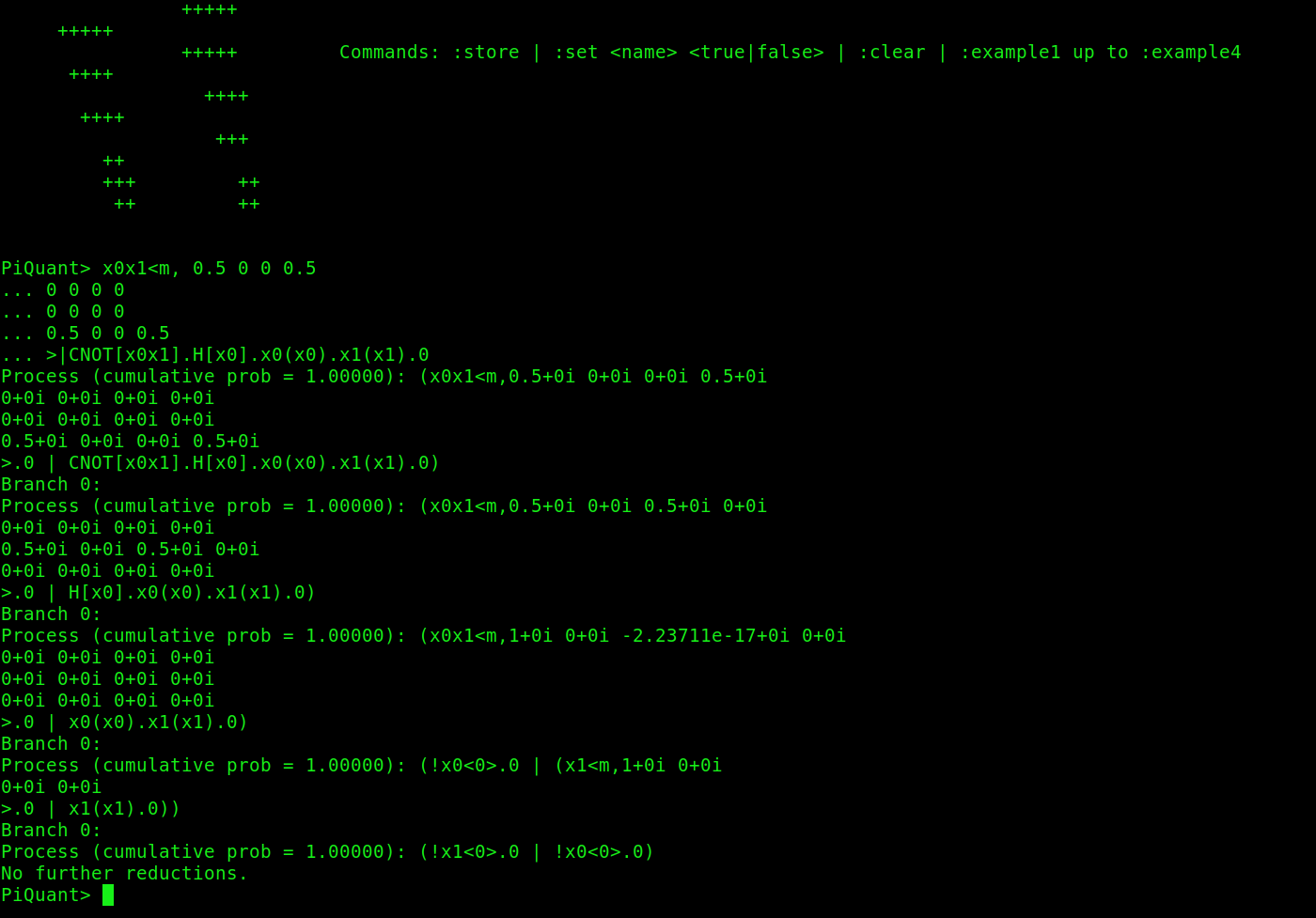}
\end{figure}

This example corresponds to \ref{bell_unentangle_ex}. Here the CNOT gate is applied to the Bell state:
\begin{equation*}
	\ket{\phi^+} = \frac{1}{\sqrt{2}} (\ket{00} + \ket{11}) = \frac{1}{\sqrt2}\begin{pmatrix}
		1 \\
		0 \\
		0 \\
		1
	\end{pmatrix}
\end{equation*}
whose corresponding density matrix is:
\begin{equation*}
	\rho = \frac{1}{2}\begin{pmatrix}
		1 &0 &0 &1 \\
		0 &0 &0 &0 \\
		0 &0 &0 &0 \\
		1 &0 &0 &1 \\
	\end{pmatrix}
\end{equation*}
Subsequently the H gate is applied (upper wire of the quantum circuit) and then measurements are performed. The result obtained is in line with the theoretical result obtained in \ref{bell_unentangle_ex}.

\begin{figure}[H]
	\includegraphics[scale=0.3]{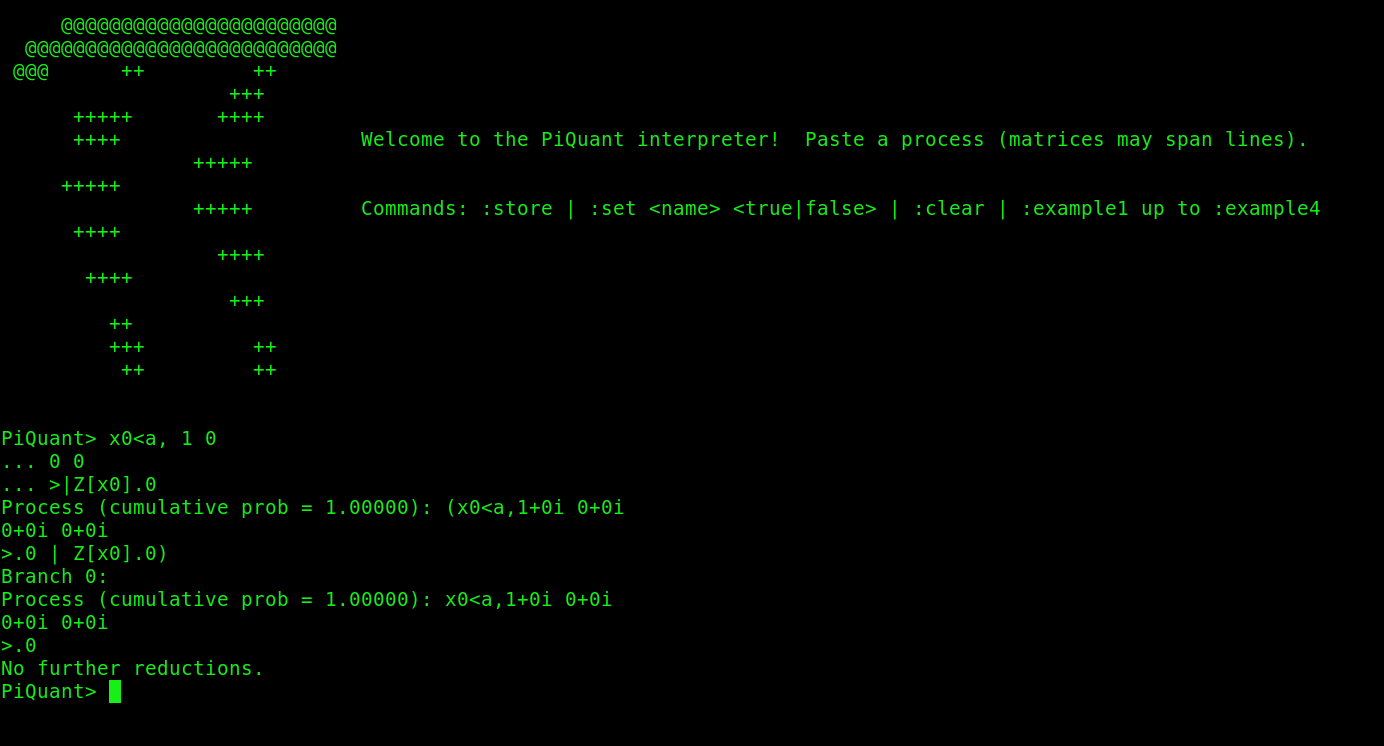}
\end{figure}

This example corresponds to \ref{z_ex}. The Z gate is applied to the state $\ket{0}$.
The result is as expected, namely:
\begin{equation*}
	Z \ket{0} = \ket{0}
\end{equation*}

\begin{figure}[H]
	\includegraphics[scale=0.3]{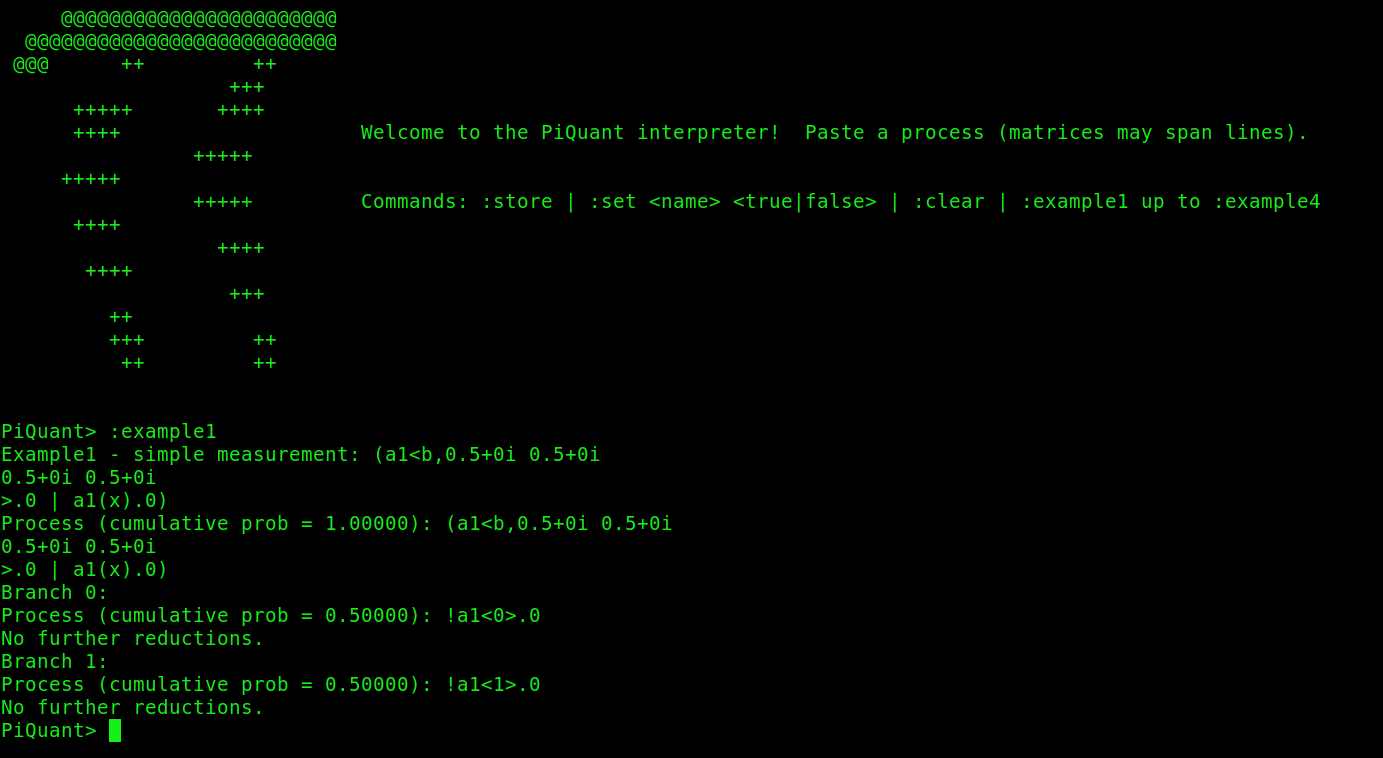}
\end{figure}

This example shows a measurement performed on the state:
\begin{equation*}
	\ket{+} = \frac{1}{\sqrt{2}} \begin{pmatrix}
		1 \\
		1
	\end{pmatrix}
\end{equation*}

As expected, the returned result is represented by 2 branches, with equal probabilities, the first one showing a measured result of 0, and the second showing a measured result of 1.

\begin{figure}[H]
	\includegraphics[scale=0.3]{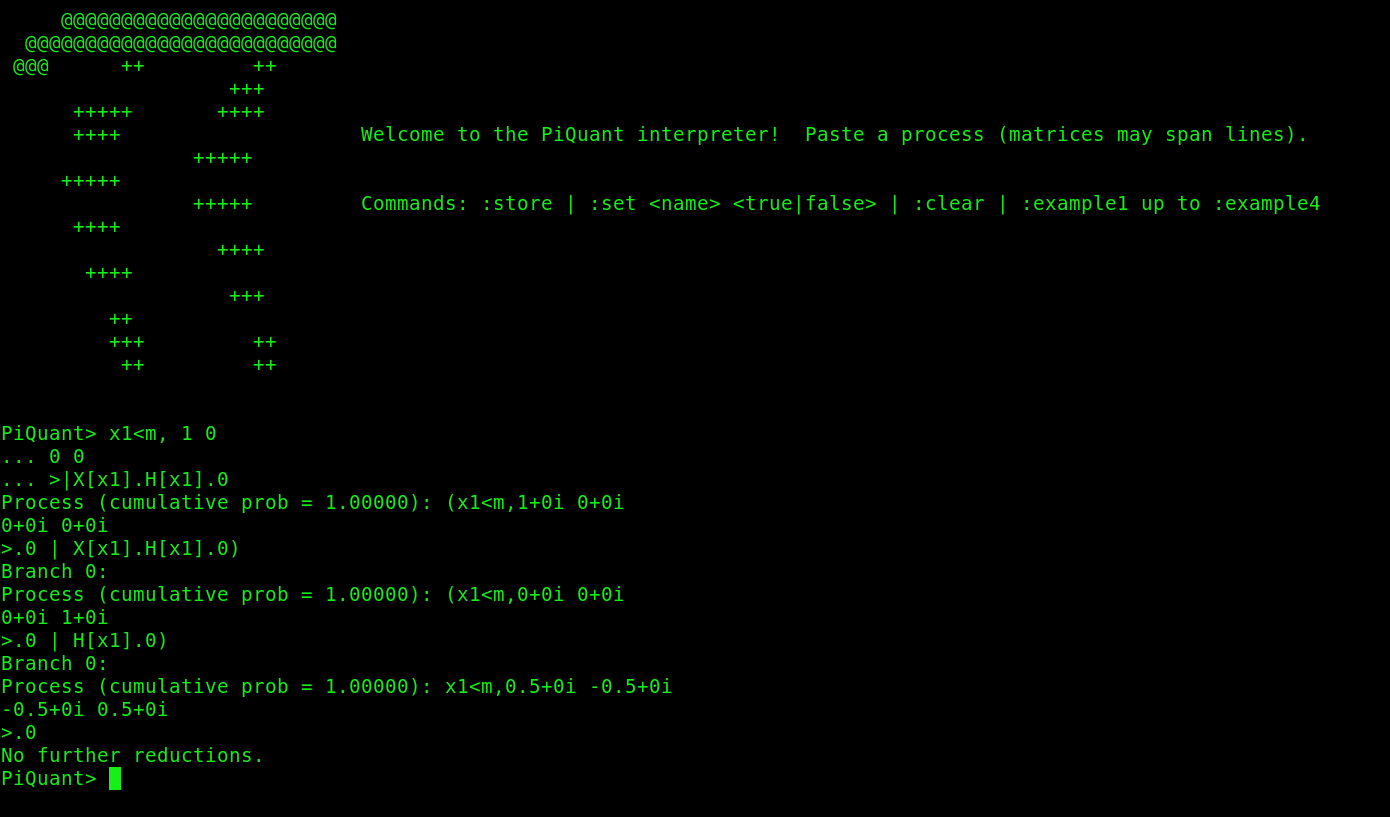}
\end{figure}

This example takes the state:
\begin{equation*}
	\ket{0} = \begin{pmatrix}
		1 \\
		0
	\end{pmatrix}
\end{equation*}
First the matrix X is applied, and then H. In quantum computing X is a bit flip operation, so the operation results in 
\begin{equation*}
	X\ket{0}= \ket{1}
\end{equation*}
The matrix H is applied next, resulting in:
\begin{equation*}
	H\ket{1}= \ket{-}=\frac{1}{\sqrt{2}}\begin{pmatrix}
		1 \\
		-1
	\end{pmatrix}
\end{equation*}
The returned result contains the density matrix corresponding to this state, which is:
\begin{equation*}
	\ket{-}\bra{-}= \frac{1}{2}\begin{pmatrix}
		1 & -1 \\
		-1 & 1
	\end{pmatrix} = \begin{pmatrix}
		0.5 & -0.5 \\
		-0.5 & 0.5
	\end{pmatrix}
\end{equation*}

\begin{figure}[H]
	\includegraphics[scale=0.3]{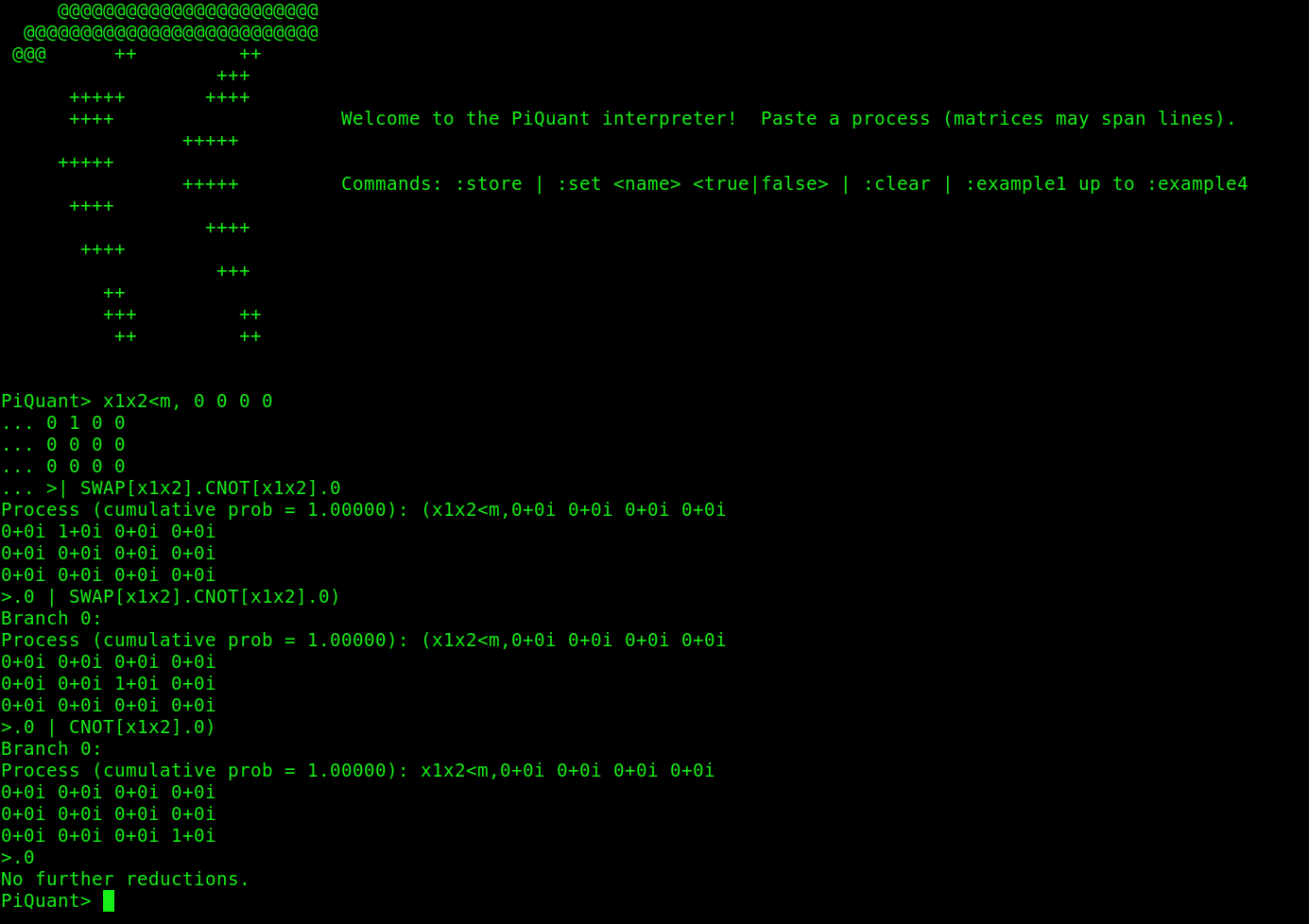}
\end{figure}

In this example, the starting point is the state:
\begin{equation*}
	\ket{01}= \begin{pmatrix}
		0 \\
		1 \\
		0 \\
		0 \\
	\end{pmatrix}
\end{equation*}
To this state the SWAP operator is applied. This changes the state to $\ket{10}$. 
Next, the CNOT operator is applied. The CNOT, or controlled NOT does the following:
\begin{equation*}
	\ket{a,b} \rightarrow \ket{a, a \oplus b}
\end{equation*}

where $a, b \in \{0,1 \}\}$. in our case, since $a=1$ and $b=0$, we expect the result to be $\ket{11}$. 
The density matrix associated to this state is:
\begin{equation*}
	\ket{11}\bra{11}= \begin{pmatrix}
		0 & 0 &0 &0 \\
		0 & 0 &0 &0 \\
		0 & 0 &0 &0 \\
		0 & 0 &0 &1 
	\end{pmatrix}
\end{equation*}

which is indeed the returned result.

\subsection{Limitations}
PiQuant has some limitations, both on the theoretical side and in its implementation.
For one, there are insufficient operations to implement circuits such as the one used to describe quantum teleportation~\cite{nielsen_chuang}, where the measured value of a qubit can determine whether a certain gate is applied in subsequent steps or not. The If operation, as currently defined in the OCaml implementation cannot evaluate the result of a measurement, only expressions involving predefined booleans. Nested If statements - i.e. passing an If statement to a "then" or "else" branch - are currently not implemented. Another limitation is the restricted way in which some arguments are passed in the OCaml implementation. for the CNOT operation, for example, only CNOT[x1x2] is accepted, CNOT[x1,x2] is not. 

Another potential issue is the fact that PiQuant primarily works with matrix multiplication and, for systems involving large numbers of qubits, the density matrices can quickly become very large. This can lead to memory bottlenecks.

\section{Conclusions}\label{sec:conc}



PiQuant is an extension of the $\pi$-calculus that models quantum-specific constructs, qubits and quantum variables. 
Qubits and multi-qubit states can both be represented --- including entangled states. 
Basic gate operations can also be performed, allowing for the simulation of a wide range of quantum algorithms. 
Measurement can also be performed, and a probabilistic choice operator is currently used to represent the possible results of measurements in a quantum circuit. 
This creates a precise semantics and understanding of the quantum states.
Whereas this does not solve completely the intuitive understanding of quantum programs, we believe it offers a better framework for grasping the concepts of superposition and collapsing states.


Multiple avenues of future research and development exist. 
Improving the existing implementation would be a first step. 
We will focus our future work on developing additional syntax and semantics rules that will allow PiQuant programs to be created in an even more intuitive manner for regular developers.


\bibliography{LambdaRhyme}
\bibliographystyle{ACM-Reference-Format}

\end{document}